# SCIENTIFIC DATA

OPEN

## Data Descriptor: A longitudinal dataset of five years of public activity in the Scratch online community

Benjamin Mako Hill[1] & Andrés Monroy-Hernández[1,2]

Scratch is a programming environment and an online community where young people can create, share, learn, and communicate. In collaboration with the Scratch Team at MIT, we created a longitudinal dataset of public activity in the Scratch online community during its first five years (2007–2012). The dataset comprises 32 tables with information on more than 1 million Scratch users, nearly 2 million Scratch projects, more than 10 million comments, more than 30 million visits to Scratch projects, and more. To help researchers understand this dataset, and to establish the validity of the data, we also include the source code of every version of the software that operated the website, as well as the software used to generate this dataset. We believe this is the largest and most comprehensive downloadable dataset of youth programming artifacts and communication.

| | |
|---|---|
| **Design Type(s)** | observation design • longitudinal data analysis • database creation objective |
| **Measurement Type(s)** | exploration behavior |
| **Technology Type(s)** | digital curation |
| **Factor Type(s)** | |
| **Sample Characteristic(s)** | Homo sapiens |

[1]Department of Communication, University of Washington, Box 353740, Seattle, Washington 98195, USA. [2]Microsoft Research, Redmond, Washington 98052, USA. Correspondence and requests for materials should be addressed to A.M.-H. (email: andresmh@uw.edu).

## Background & Summary

The Scratch online community is a public and freely accessible website where young people create, share, and interact with one another around multimedia projects, like animations and video games, created using the Scratch programming environment[1]. The dataset we present is a collection of all public data recorded by the Scratch website software over the site's first five years. The data covers the period from the website's creation in March 2007 through April 1, 2012. This includes a short period before May 14, 2007 after the site had been created but before it had been publicized widely. The dataset includes 32 tables extracted from the MySQL database used to run the Scratch website. The tables capture many types of information including metadata around users and projects, as well as user behavior and communication on the site. In addition to metadata, the dataset also includes the full text of public comments, the full source code for every project, and quantitative summaries of the contents of projects. The dataset includes data on more than 1 million users, nearly 2 million projects, more than 10 million comments, and metadata on many tens of millions of events like views, downloads, and expressions of appreciation.

Our primary goal in the release of these data is to make it easier for researchers to study how young people learn, create, communicate, and interact in informal learning environments—especially around computer programming. There are a number of public websites, including Scratch, where young people learn to program. However, even when data from these communities are available on public websites, accessing data requires writing custom software to download and parse (i.e., 'scrape') web pages[2] —actions that are frequently prohibited by websites' terms of service. Additionally, to collect granular metadata on the timing of online events, researchers need to visit pages with frequent repetition, which cannot be done retrospectively. As a result, observational research of learning online has typically been completed by computer scientists and engineers able to write and run such custom 'scraping' software and, even so, typically obtains only small samples of interaction data over small windows of time. There are several examples of such studies of Scratch (e.g., ref. 3) which, due to resulting small samples, can lead to unrepresentative data and misleading results[4].

Due to these challenges, research using long-term digital trace data from online communities of any magnitude is often done by groups with direct access to the databases that run websites. For example, the Scratch Team and its collaborators have published peer-reviewed papers, theses, and books using what has been, until now, exclusive access to 'server-side' Scratch data[5–13] (a more comprehensive list can be found at https://scratch.mit.edu/info/research/). Our goal is to increase the number of researchers who can engage in this work. Also, by providing access to the full dataset, we hope to reduce the issues that arise from using small data samples of Scratch activity. We hope that the release of these data will be used to develop and test theories in diverse fields of study including communication, the learning sciences, computer science, the social sciences, and digital humanities research.

## Methods

This dataset describes activities of participants in the Scratch online community. Activities represented in the dataset include viewing, creating, and sharing interactive computer programs, referred to as 'projects,' as well as interacting with other participants. Participants created projects using a desktop program (called the *Scratch Authoring Environment* or Scratch for short) which could then be uploaded to the Scratch online community website. In this section, we attempt to provide context for this dataset by briefly describing how participants created and shared their projects and interacted with one another, before we explain the process we followed to construct the dataset.

### Setting

Scratch is a project of the Lifelong Kindergarten Group at the MIT Media Lab that seeks to help young people learn to 'think creatively, reason systematically, and work collaboratively'[1]. Using the Scratch Authoring Environment, people can program their own interactive stories, games, and animations, and share those creations with others on the Scratch Online Community website. Similar to Logo, its intellectual predecessor, Scratch was inspired by pedagogy that aims to enable learning by creating personally meaningful programs, often with or around others[14].

### Authoring environment

People use the Scratch Authoring Environment to create 'projects' by constructing on-screen objects called 'sprites' containing images, sound, and computer code. Code consists of visual blocks (similar to tokens in other programming languages) that creators place together like puzzle pieces to form groups called 'scripts' that control the appearance and behavior of sprites. There are more than 120 unique blocks with differing functionality, colors, and shapes.

Users create scripts to accomplish a wide range of tasks. For example, scripts might change the image representing a sprite (i.e., 'costume'), its location on the screen, or its size, color, or sounds. Scripts can also control how sprites interact with the user or with other sprites. Sprites might also be programmed to take input dynamically through, for example, a computer keyboard, a mouse, or a microphone.

The Scratch Team released five desktop-only versions of Scratch from 2007 to 2009: versions 1.0 through 1.4 (1.x for short). A web-based version, known as Scratch 2.0, was released in 2013. This data set only contains projects created with Scratch 1.x. Users were able to download Scratch 1.x (pictured in

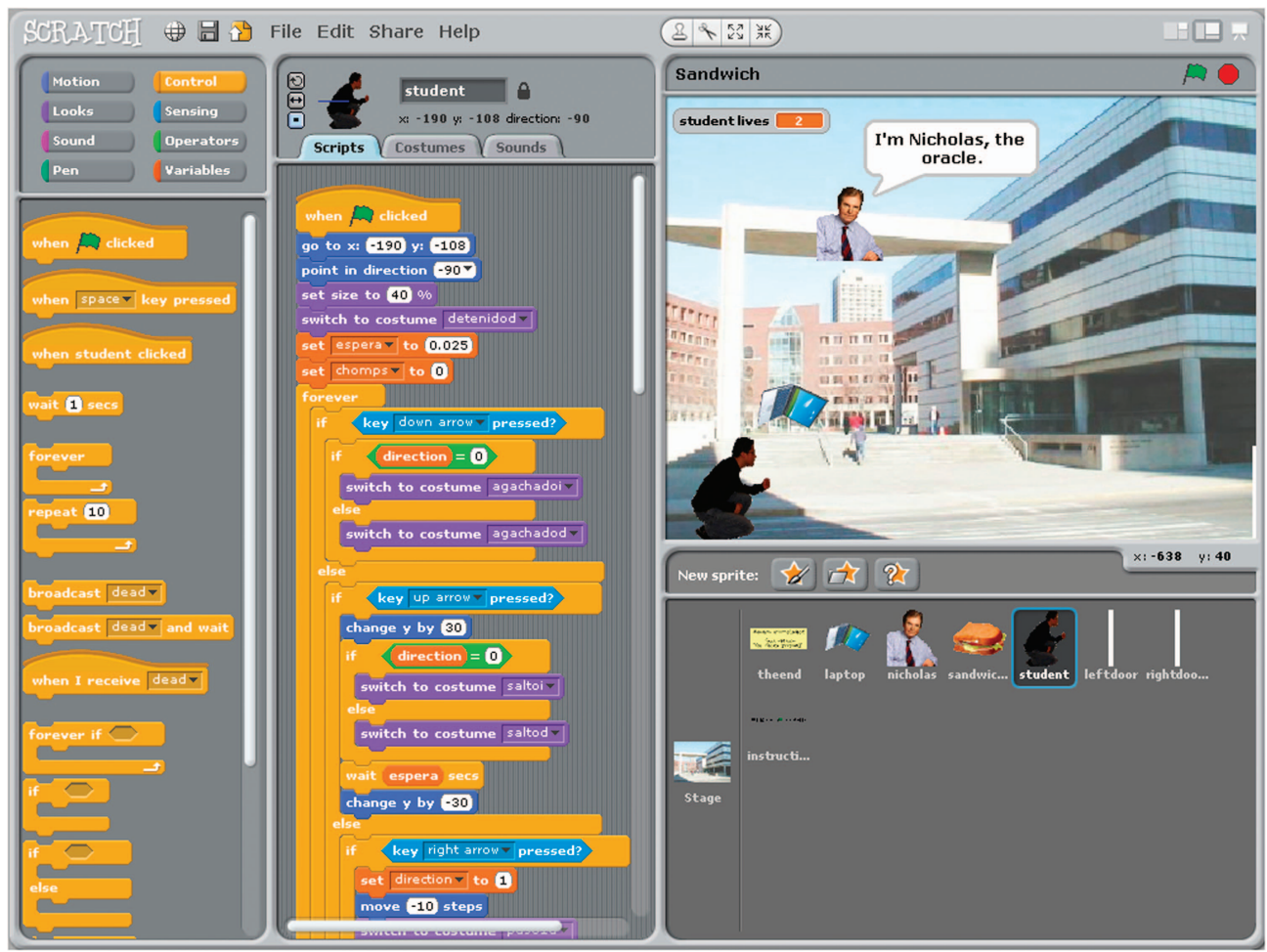

**Figure 1. Screenshot of the *Scratch Authoring Environment*.** A collection of blocks is shown in the left-most column. In the center column, blocks are assembled together in one large script. The bottom right shows a list of sprites. The top right shows the platform in which the full program will run.

Fig. 1) onto their Windows, Mac OS X, or Linux computers. The process of installation usually required people to download the software from the Scratch website. In some cases, like in the XO machine by One Laptop per Child, Scratch 1.x was pre-installed. These 1.x versions of Scratch were created using Squeak, a dialect of the Smalltalk programming language.

### Online community

The Scratch website was conceived as an online space where Scratch creators could access an audience, find potential collaborators, and explore a repository of inspirational creations to study and learn from[15]. In particular, the site sought to engage youth through remixing—the creative recombination of material created by others. The website drew ideas from previous research on constructionist learning communities (e.g., ref. 16) and commercial social network sites. Projects shared on the Scratch website vary in complexity and style, ranging from animations to simulations, video games, interactive stories, and art and music.

Users could upload their projects to the website by clicking the 'Share' button inside the Authoring Environment. When sharing projects, users were prompted to enter a name for their project, a description, categorization tags, their account's username, and a password for the online community. Users without accounts were required to create them before sharing their projects. After being shared, projects were publicly available on the website so other people could view the project, interact with it, download it, remix it, comment on it, and more. Although visiting and running projects did not require an account, most 'active' forms of engagement, like commenting, did.

In addition to storing metadata about projects, users, and interactions, the website stored each of the raw Scratch project files which were automatically parsed and analyzed by the website software to extract summary information about project content, including programming code and media elements. A community-built parser for Scratch 1.x projects is publicly available at https://github.com/tjvr/kurt. The Scratch online community website uses a custom-built database-driven web application. The software running the website from 2007 to 2012 was named *ScratchR* (short for Scratch Repository) and was written in PHP and MySQL[15]. For the release of Scratch 2.0 in 2013, the Scratch Team transitioned to a version of the website with a new underlying architecture.

### Defining public data

To respect the privacy of users of the Scratch website, this dataset is restricted to data made public by users of the site with the knowledge that it would be fully visible online. The heuristic we used to determine if content was public in this sense was if the data would have been visible to a piece of web-scraping software visiting the Scratch site frequently *and* if no subsequent action by a Scratch user or administrator would have hidden the data. We include every piece of data in the Scratch database that adheres to a conservative application of this heuristic while consistently seeking to minimize potential risk to human subjects. For example, we include detailed timestamp data for events that occurred publicly, like the time that the download count for each project was incremented. Although precise logs of these events were not systematically published after they occurred, these events were made public as the happened through counters on the site and would have been visible to a web archiver or scraper.

Some data are excluded because they were never public. For example, we omit self-reported user age, gender, email addresses, IP addresses, and reports of inappropriate content made to administrators. Similarly, information on which users participated in public interactions where a participant in the interaction was not made public (e.g., which user downloaded a project) are omitted.

We also omit any data removed by users, even though the data were public at some point in the past. For example, we omit all information about projects that were removed or made 'invisible' by their authors. Because users may have updated projects with new versions to hide detail about their creations, we also omit data about previous versions of projects. We also omit all activity that was publicly attributed to users who had deactivated their accounts at the point that we collected these data.

We built this dataset using the final version of the database at the point that the *ScratchR* website was shut down on May 6, 2013 to transition to Scratch 2.0. Because our most recent records are from April 2012, participants each had more than one year to remove their data before this dataset was produced.

### Procedures

We began the process of creating this dataset by reviewing every field of every table of each of the two MySQL databases used in the live Scratch website to determine if the data were public according to the definition described above. One database, called 'beta,' was used by the website to generate pages. Another, called 'project_analytics,' was populated with information extracted from projects as they were uploaded. We described all fields in a series of comma separated values (CSV) codebooks that we have included in the dataset. Furthermore, we looked for examples of observations or rows that were not public and omitted these as well.

### Code availability

We wrote custom GNU R programs to export and process each table. Each R program includes an SQL statement that extracts the full MySQL table. We executed these SQL statements within R using the RMySQL software package (https://cran.r-project.org/web/packages/RMySQL/). Using R, we also loaded our CSV codebooks and dropped all non-public columns. In the same R programs, we recoded and 'cleaned' variables and dropped any rows that needed to be omitted because they included non-public data. Finally, the processed data was exported directly from R into the final file formats. Every omitted field is noted in the codebooks provided with the release and every omitted observation is reported in the documentation provided with the datasets. The full source code of each R program is provided with the dataset (Data Citation 1).

### Research ethics

The users of the Scratch online community are young people, many under 18 years of age, from countries across the world. In designing the release of these data, we have worked to insure that the data collected here contains no private data. The Scratch website rules prohibit users from sharing identifiable information in any form and even discourages usernames that might hint at 'real' names or identities. Even so, and even though these data are public and were contributed by users who knew they would be public, there may still be risks involved in research using these data. As a result, the Scratch Team at MIT is requiring that all researchers seeking access to these data provide their name, affiliation, and a description of their project so that it can be reviewed by the Scratch Team first. Moreover, they are requiring that researchers using these data accept a data usage agreement described in the Usage Notes section below.

The protocol for the publication of these data was approved by the Committee on the Use of Humans as Experimental Subjects at MIT. Researchers using these data are responsible for ensuring that their own use and research using these data is conducted with any necessary approval from their own ethics or institutional review boards.

### Data Records

The dataset is made available on a website hosted by the Lifelong Kindergarten group (LLK) at the MIT Media Lab https://llk.media.mit.edu/scratch-data/ with an archival copy placed in the Harvard Dataverse network (Data Citation 1). The website includes a web page for each table which contains: an expanded version of the summary information below describing the type of data stored in each table and additional context and explanation; the content of each codebook including a description of every field in the table, the type of data stored in the field (i.e., integer, character string, etc.); and basic summary statistics, tables, and visualizations (e.g., visualizations of the distribution of observations in a table over time). Critically, the page also includes an accounting of every time observations were omitted from the dataset. Finally, each page includes hyperlinks to the downloadable datasets.

The datasets are included in three different formats. All datasets are included in 'RDATA' format. RDATA is an external representation of objects in the GNU R programming language. The objects can be loaded into R using the function 'load()' or 'attach()'. RDATA is an implementation of External Data Representation (XDR) which is a standard data serialization format. Numeric data are also provided in comma separated values format (CSV). Textual data are provided in in JavaScript Object Notation (JSON) format. Because the JSON files are very large, they are not formatted as a single JSON object. Instead, they are formatted as a 'stream' of JSON dictionaries where each observation is formatted as a separate JSON dictionary, separated by newlines. Newlines and carriage returns ('\ n' and '\ r') within text fields are escaped. This format is very similar to the format used for other social media data including data returned by the Twitter streaming API and GitHub event data.

The dataset is separated into three sections: core tables, text tables, and project analytics tables. Each of the tables is described briefly below. A summary of the number of observations, fields, and formats are in Table 1.

### Core tables

Core tables contain data and metadata documenting the major objects and relationships captured by the Scratch online community website. The data in these tables are numeric and do not include user-inputted text, which are stored in text tables. These datasets are provided in RDATA format and comma-separated values format (CSV). This portion of the dataset includes the following 18 tables:

**users** Each row in this table represents a user account that was publicly visible on the Scratch website at the time that these data were collected. Each account had a public profile web page (see Fig. 2).

**projects** Each row in this table represents a project that was publicly visible on the Scratch website at the time that these data were collected. Projects were the interactive creations (e.g., video games, interactive art, simulations, and animations) that users shared onto the website. Each project had its own web page (see Fig. 3).

**galleries** Each row in this table represents a gallery that was publicly visible on the Scratch website at the time that these data were collected. Galleries were collections of projects displayed on a web page (see Fig. 4).

**friends** Each row in this table represents the event of a user 'friending' another user. The 'friendships' reflected in this dataset include all friendships that were publicly visible on the Scratch website at that time that these data were collected. Friendships were unidirectional relationships between users displayed in the website as 'friends.'

**downloaders** Each row in this table a represents the event of a user downloading a project. In order to offer the most consistent measure of downloads, this dataset counts only the first download per user. The identity of the user downloading a project was not publicly displayed, so it is not included in this table.

**favoriters** Each row in this table represents the event of a user adding a project to their favorites (i.e., bookmarking it). The list of favorites appeared in the user's profile page.

**lovers** Each row in this table represents the event of a user clicking a heart-shaped 'love-it' button that appears on every project's page. This action was socially framed as an expression of appreciation for a project. The identity of the user 'loving' a project was not publicly displayed and is not included.

**viewers** Each row in this table represents the event of a user viewing or loading the webpage of a project for the first time. The identity of the user viewing a project was not publicly displayed and is not included.

**project_comments** Each row in this table represents a comment left on a project. This table includes all comments that were publicly visible on every project page in the Scratch website at the time of composing this dataset. This table contains the metadata for these comments and includes who, when, and on which project each comment was posted. The text or content of each comment is in the **pcomments_text** text table.

**gallery_comments** Each row in this table represents a comment left on a gallery. This table includes all comments that were publicly visible on every gallery page in the Scratch website at the time that these data were collected. This table contains the metadata for these comments and includes who, when, and in which gallery each comment was posted. The text or content of the comment is in the **gcomments_text** text table.

**projects_galleries** Each row in this table reflects a project's inclusion in a gallery at the point of data collection. Projects are listed multiple times in this table if they were included in multiple galleries.

| | Table Name | Observations | Fields | File Formats |
|---|---|---|---|---|
| 1 | users | 1,056,951 | 4 | CSV, RDATA |
| 2 | projects | 1,928,699 | 44 | CSV, RDATA |
| 3 | galleries | 120,097 | 3 | CSV, RDATA |
| 4 | friends | 1,313,200 | 3 | CSV, RDATA |
| 5 | downloaders | 4,236,575 | 2 | CSV, RDATA |
| 6 | favoriters | 1,041,387 | 3 | CSV, RDATA |
| 7 | lovers | 1,487,112 | 2 | CSV, RDATA |
| 8 | viewers | 30,154,002 | 2 | CSV, RDATA |
| 9 | pcomments | 7,788,414 | 5 | CSV, RDATA |
| 10 | gcomments | 2,817,973 | 5 | CSV, RDATA |
| 11 | projects_galleries | 908,717 | 3 | CSV, RDATA |
| 12 | tags_projects | 2,905,774 | 3 | CSV, RDATA |
| 13 | tags_galleries | 222,064 | 3 | CSV, RDATA |
| 14 | frontpage_projects | 32,208 | 3 | CSV, RDATA |
| 15 | featured_projects | 1,227 | 2 | CSV, RDATA |
| 16 | featured_galleries | 252 | 2 | CSV, RDATA |
| 17 | studio_galleries | 61 | 2 | CSV, RDATA |
| 18 | curators | 72 | 2 | CSV, RDATA |
| 19 | projects_text | 1,928,699 | 3 | JSON, RDATA |
| 20 | galleries_text | 120,097 | 3 | JSON, RDATA |
| 21 | pcomments_text | 7,788,414 | 2 | JSON, RDATA |
| 22 | gcomments_text | 2,817,973 | 2 | JSON, RDATA |
| 23 | tags_text | 353,934 | 3 | JSON, RDATA |
| 24 | project_block_stacks | 13,219,765 | 4 | JSON, RDATA |
| 25 | project_block_stacks_disconnected | 2,774,865 | 3 | JSON, RDATA |
| 26 | project_strings | 44,789,099 | 4 | JSON, RDATA |
| 27 | project_blocks | 1,925,606 | 172 | CSV, RDATA |
| 28 | project_drums | 259,663 | 3 | CSV, RDATA |
| 29 | project_media | 44,205,193 | 5 | CSV, JSON, RDATA |
| 30 | project_midi_instruments | 128,571 | 3 | CSV, RDATA |
| 31 | project_save_history | 12,590,095 | 2 | CSV, RDATA |
| 32 | project_sprites | 12,987,630 | 4 | CSV, RDATA |

Table 1. Overview of each of the tables included in the dataset including the name of the table, the number of observations included in the databases, the number of fields or columns, and the file formats provided.

   **tags_projects** Each row in this table represents a tag placed on a project. Projects can have zero or more tags. Each tag could also be associated with additional projects or galleries. Tags were publicly displayed on project pages.
   **tags_galleries** Each row in this table represents a tag placed on a gallery. It is similar in structure and purpose to the **tags_projects** table. Tags were publicly displayed on gallery pages. Like projects, galleries could have zero or more tags and each tag could be associated with additional projects and galleries.
   **frontpage_projects** Each row in this table represents a project that was displayed on the front page for any of several reasons (see Fig. 5). For example, projects could be placed on the front page automatically for being among the most remixed or the most viewed projects on the website. They could also be added manually by an administrator or a curator.
   **featured_projects** Each row in this table represents a project that was displayed on the front page of the Scratch website in a section called 'Featured Projects' that held three projects at a time (see Fig. 5). Only Scratch website administrators could add projects to this section. The process of selecting featured projects was entirely manual and was usually driven by a decision about the quality and interestingness of the project.
   **featured_galleries** Each row in this table represents a gallery that was displayed on the front page of the Scratch website in a section called 'Featured Galleries.' Featured Galleries worked identically to Featured Projects and were displayed in the same way.

**Figure 2. Example profile page for the user 'SampleProjectsTeam' on the Scratch website.** By default, the page shows a user's most recent 15 projects and includes links to additional pages with older projects. Because the user has not 'friended' (i.e., followed) any others, there is a box in the left-most column explaining that they have 'No friends yet.'.

Figure 3. **Example of a page created for a project on the Scratch website.** Each project had a similar page associated with it. Visitors to the page can view and interact with the project in the upper right half of the page. The rest of the page is dedicated to presenting project metadata with a space for commenting below.

**studio_galleries** Each row in this table represents a gallery that was displayed on the front page in a section called 'Design Studio.' Only the website administrators could add galleries to this section. The process of adding new galleries was manual and the featured galleries were usually created by administrators.

**curators** Each row in this table represents a user who at some point was in charge of selecting projects for the front page of the website, in a section labeled 'Curated By' (see Fig. 5). This section displayed three

**Figure 4. Example of gallery page on the Scratch website.** Like the user profile page shown in Fig. 2, galleries primarily include lists of projects listed in reverse order of when they were added to the gallery. A description and basic metadata are shown in the left column. Not shown in the figure, there is a space for comments on the gallery itself below the list of projects.

..................................................................................................................................

projects selected by the curator. When the curator added a project to their list of favorites, the project would be displayed in the 'Curated By' section automatically. There was only one active curator at a time.

**Text tables**
The 'text' tables include textual data created by users. In each case, these tables correspond to a table included among the core tables; each includes ID numbers corresponding to observations in those tables. We have separated these tables both because these data are very large and because there are challenges specific to encoding and escaping user-inputted data making it impossible to produce CSV files that can be reliably loaded across statistical software packages and spreadsheet applications.

**Figure 5. Snapshot of the front page of the Scratch website that was presented to users at http://scratch.mit.edu/.** The top row of projects are projects 'featured' by administrators. The second row includes projects selected by a 'curator' whose username is 'the_hawk_arisen'. The header helped users navigate to different sections of the website.

Some of these tables include data with unknown or invalid text encoding. The Scratch website was designed to only record UTF-8 encoded Unicode text. However, errors and data corruption meant that some of the data in the MySQL database are invalid UTF-8 text. In general, we have elected to include these poorly encoded data because they are the actual data submitted by users and displayed on the website. These datasets are all provided in RDATA format and as streams of JSON dictionaries. We have included software in Python for loading JSON files that will identify and truncate invalid text records. The text portion of the dataset includes the following 8 tables:

**projects_text** Each row in this table represents a project that was publicly visible on the Scratch website at the time that these data were collected. This table contains the free-form and the unstructured text fields in the projects table.

**galleries_text** Each row in this table represents a gallery. This table contains the free-form and the unstructured text fields in the galleries table.

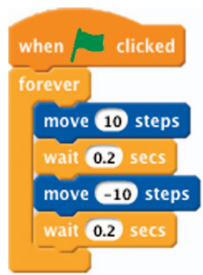

**Figure 6. Example of a stack of blocks. In Scratch, the execution of code is triggered by a 'hat' block that determines when the stack gets executed.** For example, the execution of the blocks in the image shown is triggered by the 'hat' block 'when [green flag] clicked'.

**comments_text** Each row in this table represents a comment left on a project. This table contains the free-form and the unstructured text fields in the **pcomments** table.

**gcomments_text** Each row in this table represents a comment left on a gallery. This table contains the free-form and the unstructured text fields in the **gcomments** table.

**tags_text** Each row in this table represents a tag used on a project and/or gallery. This table contains the free-form and the unstructured text fields in the tables **tags_projects** and **tags_galleries**.

**project_block_stacks** Each row in this table represents the textual representation of a code block associated with a sprite used in the most recent version of a project shared on the Scratch website at the point of data collection. The code is presented in both a human-readable form, and in the raw format as it is stored in the Squeak programming language (used to create this version of Scratch). The table only includes block stacks that are intended to be executed (see Fig. 6 for an example).

**project_block_stacks_disconnected** Each row in this table represents Scratch blocks that are never executed (i.e., those blocks that do not have an execution trigger or 'hat block' on top).

**project_strings** Each row in this table represents a text string—free-form text strings typed by users as part of their code—used in the most recent version of a project shared on the Scratch website at the point of data collection. These text strings include variable names, and messages that are printed on the screen.

**Project analytics tables**

Although the Scratch website primarily recorded data about interactions around projects, the website software parsed projects to store basic details about the contents of all project files. This included information on project metadata like the number and type of blocks used, the number and type of media elements used, and the history of when projects were saved. The project analytics tables include these metadata. These data are missing for several thousand project files which the website software could not parse. A examination of several of these projects suggests that many of these projects were corrupt or created with modified versions of the Scratch Authoring Environment software that resulted in non-standard binaries.

These datasets are provided in RDATA format and in comma separated values (CSV) format. The **project_media** table also contains user-inputted text strings corresponding to filenames for media files and, as a result, is provided in JSON as well as CSV and RDATA. This portion of the dataset includes the following 6 tables:

**project_blocks** Each row in this table represents a project shared on the Scratch website at the point of data collection. Each column of this row represents the frequency counts of each block type for a particular project. The value of each cell represents the number of times a particular block is used.

**project_drums** Each row in this table represents a drum used in the most recent version of a project shared on the Scratch website at the point of data collection. The list of drums was taken from the General MIDI Level 1 Percussion Key Map.

**project_media** Each row in this table represents a media item (e.g., an image or a sound) attached to a sprite in the most recent version of a project shared on the Scratch website at the point of data collection. A single project could have many pieces of media.

**project_midi_instruments** Each row in this table represents a musical instrument used in the most recent version of a project shared on the Scratch website at the point of data collection. Instrument types included recognizable instruments like pianos, guitars, and flutes in the General MIDI Level 1 Instrument Patch Map.

**project_save_history** Each row in this table represents a time when a user saved a project to their local storage device (e.g., hard drive) in the most recent version of a project shared publicly on the Scratch website at the point of data collection. This data was generated from a log of save 'events' that was included in each project. If a project was a remix, its log included the full log of the project (or projects) on which the current project was based.

**project_sprites** Each row in this table represents a sprite used in the most recent version of a project shared on the Scratch website at the point of data collection. The table includes information on the number of scripts, sounds, and images associated with each sprite.

## Technical Validation

We identified two distinct senses in which the validity of these data might be assessed. In the first sense, these data would be said to be valid to the extent that they are a reliable account of user behavior in the Scratch online community in the period that the dataset covers. In the second more limited sense, these data would be said to be valid if they are an accurate representation of the data displayed to the public, and to users of the Scratch website, in December 2012.

Assessing the degree to which these data are a valid reflections of user activity between 2007 and 2012 is extremely difficult because this database represents the only record of this activity. Additionally, we know that these data are an incomplete accounting in several ways. For example, we have intentionally omitted data that is not public.

There are also unintentional ways in which this dataset's accounting of activity may be incomplete or incorrect. For example, there were about 3,000 duplicated friend relationships out of a total of 1.3 million caused by a software glitch. The Scratch online community was under active development and active use by more than a million people for a period of five years, and there is evidence that errors occurred that impacted data collection. For example, in ways that we detail in the project documentation, our project analytic code found examples of data corruption and duplicated data records. We cannot know if, or how many, records of interaction are missing or wrong.

With these important limitations in mind, there are also reasons to believe that the data are a reliable representation of user behavior. The best evidence we can provide to support this claim is a complete accounting of the software used to record the data originally. To this end, we have prepared an archive of the software used to create these data over the entire period covered in the dataset and we have included this software archive in the dataset. First, we have created a bundle of the full source code of *ScratchR*. We converted the revision control repository for the software (originally Subversion) into Git and included the repository with full history alongside the dataset. The repository contains 1,637 distinct versions of the software. This means that researchers can see every version of the software and can retrieve particular versions as they existed at precise points in time. Second, we have created a bundle that includes the full source code to the Scratch Authoring Environment. We include code for the three versions that were in active use during the period that the dataset was collected (versions 1.2, 1.3, and 1.4). *ScratchR* and the Scratch Authoring Environment are made available under the GNU General Public License Version 2 and the Scratch Source Code License. We have also included binary copies of the Scratch Authoring Environment version 1.4 for Microsoft Windows, Mac OSX and Debian or Ubuntu versions of GNU/Linux.

In the second sense that these data would be considered valid if they are accurate reflections of data displayed to the public and to users on the Scratch website in December 2012, we can have much more confidence. To this end, we have released the full source code we used to prepare this dataset from the MySQL database as described in the Procedures section. We have also provided full codebooks that detail every database field we have exported as well as every field we have not included. We have included commented code that details every step where a field is dropped, where a new variable is created, or where rows are dropped.

In the case of data that are removed, we have documented and described each of these steps in dynamically generated *knitr* documents[17] to ensure that our documentation detailing omitted data is produced automatically as part of the dataset preparation process and is fully reproducible by anybody with access to the MySQL database. Although we cannot provide access to the original database (because it contains private human subject data), the source code we have provided should detail the entirety of the data processing we have conducted and allow researchers to audit our code.

## Usage Notes

The dataset is accessible online at https://llk.media.mit.edu/scratch-data/. Users who wish to download the data should visit that URL where they should fill out a five-item form indicating agreement with the Scratch Research Data Sharing Agreement (SRDSA) and providing details on each researcher's full name, email address, affiliation, title, and a description of the project they plan to work on with the data. Once submitted, applications are processed and approved by the Scratch Team who then send an email to applicants with a URL and password that can be used to download the complete dataset over an encrypted HTTPS connection. Because it is possible that the URL above will become unavailable in the future, an archival copy of the complete dataset and documentation has been placed in the Harvard Dataverse (Data Citation 1). Although access to the data files in the Dataverse is restricted by default, the Dataverse page contains details on how to agree to the SRDSA and request access to the data. The Dataverse page will be updated if details of this process change.

The SRDSA requires that researchers use the data only for scholarly and research purposes; prohibits sharing, redistribution, or republishing of the data; prohibits attempts to identify or contact individual Scratch users; and requires that researchers attempt to maintain the anonymity of individual users. For example, although usernames are intentionally omitted, the agreement prohibits attempts to discover them using the dataset and web searches.

The agreement also asks users to not republish or quote from the data in a way that would allow readers to identify the authors of the original material. For example, strings reported in web search results (e.g., strings in the **project_text** and **galleries_text** tables) should never be quoted or included verbatim in publications. The text of comments (i.e., text stored in the **pcomments_text** and **gcomments_text**) tables, which are not included in search results, can be quoted. Finally, the agreement also requests that researchers cite both a paper describing the Scratch project[1] as well as this dataset descriptor in any publications that result from use of the dataset.

The agreement also makes two non-binding requests to researchers who use the dataset. First it requests that researchers share any code necessary to fully reproduce their results upon publication of papers. Second, it requests that users send the Scratch Team bibliographic details of any papers published so that the work can be added to a list of Scratch research the Scratch Team maintains. In both cases, emails with this information can be sent to scratchdata@media.mit.edu.

### Data Citation

### Acknowledgements

This work was supported by the National Science Foundation (grant DRL-1417663). We thank the Lifelong Kindergarten Group at the MIT Media Lab for creating Scratch, creating and continuing to support the Scratch online community over many years, and generously allowing us to help share their data with other researchers. In particular, thanks to Mitchel Resnick, Natalie Rusk, and Sayamindu Dasgupta for helping us plan the details of this release and for feedback on this manuscript.

We also deeply appreciate the help of everybody who helped test these data since an initial 'beta' release that began in 2014. In particular, we appreciate feedback from Sayamindu Dasgupta and Benjamin Berg at MIT, and Jeff Nickerson and his lab at the Stevens Institute of Technology.

### Author Contributions

B.M.H. contributed to the planning of the project, the creation of software to prepare the datasets, the creation and editing of the documentation, and the writing of the paper. A.M.H. led the design and development of the Scratch online community website that generated these data and contributed to the planning of the project, the creation of software to prepare the datasets, the creation and editing of the documentation, and the writing of the paper.

### Additional information

**Competing financial interests**: The authors declare no competing financial interests.

**How to cite this article**: Hill, B. M. & Monroy-Hernández, A. A longitudinal dataset of five years of public activity in the Scratch online community. *Sci. Data* 4:170002 doi: 10.1038/sdata.2017.2 (2017).